\documentclass[12pt]{article}

\usepackage{graphicx}

\newcommand{\bdv}[1]{\mbox{\boldmath$#1$}}

\def\rel{{\rm rel}}
\def\e{{\rm E}}
\def\au{{\rm AU}} 
\def\muas{{\mu\rm as}}
\def\kms{{\rm km}\,{\rm s}^{-1}}

\def\rel{{\rm rel}}

\def\mas{{\rm mas}}
\def\masyr{{\rm mas\,yr^{-1}}}

\def\e{{\rm E}}
\def\bpi{{\bdv{\pi}}}
\def\bgamma{{\bdv{\gamma}}}

\def\bmu{{\bdv{\mu}}}

\def\la{{<\atop\sim}}
\def\ga{{>\atop\sim}}

\def\apj{{\it Astrophys. J. \ }}
\def\aap{{\it Astron. Astrophys. \ }}
\def\mnras{{\it Mon. Not. R. Astron. Soc. \ }}

\usepackage{scicite}

\usepackage{times}

\newenvironment{sciabstract}{%
\begin{quote} \bf}
{\end{quote}}

\newcounter{lastnote}

\title{A Terrestrial Planet in a $\sim 1\,\au$ Orbit
Around One Member of a $\sim 15\,\au$ Binary} 

\author
{A.~Gould,$^{1}$, 
A.~Udalski$^{2}$,
I.-G.~Shin$^{3}$,
I.~Porritt$^{4}$,
J.~Skowron$^{2}$,\\
C.~Han$^{3,5}$,
J.~C.~Yee$^{1,6}$,
S.~Koz{\l}owski$^{2}$,
J.-Y.~Choi$^{3}$,\\
and \\
R.~Poleski$^{1,2}$,
{\L}.~Wyrzykowski$^{2,7}$,
K.~Ulaczyk$^{2}$,
P.~Pietrukowicz$^{2}$,
P.~Mr{\'o}z$^{2}$,\\
M.K.~Szyma{\'n}ski$^{2}$,
M.~Kubiak$^{2}$,
I.~Soszy{\'n}ski$^{2}$,
G.~Pietrzy{\'n}ski$^{1,8}$ \\
(OGLE Team), \\
B.S.~Gaudi$^{1}$,
G.W.~Christie$^{9}$,
J.~Drummond$^{10}$,
J.~McCormick$^{11}$,\\
T.~Natusch$^{10,12}$,
H.~Ngan$^{10}$,
T.-G.~Tan$^{13}$,
M.~Albrow$^{14}$,\\
D.L.~DePoy$^{15}$,
K.-H.~Hwang$^{3}$,
Y.K.~Jung$^{3}$,
C.-U.~Lee$^{16}$,\\
H. Park$^{3}$,
R.W.~Pogge$^{1}$,\\
($\mu$FUN Team), \\
F.~Abe$^{17}$,
D.~P.~Bennett$^{18}$,
I.~A.~Bond$^{19}$,
C.~S.~Botzler$^{20}$,
M.~Freeman$^{20}$, \\
A.~Fukui$^{21}$, 
D.~Fukunaga$^{17}$,
Y.~Itow$^{17}$,
N.~Koshimoto$^{22}$,
P.~Larsen$^{20,29}$, \\
C.~H.~Ling$^{19}$, 
K.~Masuda$^{17}$, 
Y.~Matsubara$^{17}$,
Y.~Muraki$^{17}$,
S.~Namba$^{22}$, \\
K.~Ohnishi$^{23}$, 
L.~Philpott$^{30}$,
N.~J.~Rattenbury$^{20}$,
To.~Saito$^{24}$,
D.~J.~Sullivan$^{25}$,\\
T.~Sumi$^{22}$,
D.~Suzuki$^{22}$,
P.~J.~Tristram$^{26}$,
N.~Tsurumi$^{17}$, 
K.~Wada$^{22}$, \\
N.~Yamai$^{27}$, 
P.~C.~M.~Yock$^{20}$,
A.~Yonehara$^{27}$, \\
(MOA Team), \\
Y.~Shvartzvald$^{28}$,
D.~Maoz$^{28}$,
S.~Kaspi$^{28}$,
M.~Friedmann$^{28}$,\\
(Wise Team), \\
}

\date{}

\begin{document} 

\maketitle 

\noindent
\normalsize{$^1$Department of Astronomy, Ohio State University, 140 W. 18th Ave., Columbus, OH  43210, USA} \\
\normalsize{$^2$Warsaw University Observatory, Al.~Ujazdowskie~4, 00-478~Warszawa, Poland} \\ 
\normalsize{$^3$Department of Physics, Chungbuk National University, Cheongju 371-763, Republic of Korea} \\
\normalsize{$^4$Turitea Observatory, Palmerston North, New Zealand} \\
\normalsize{$^5$Corresponding Author} \\
\normalsize{$^6$Harvard-Smithsonian Center for Astrophysics, 60 Garden St., Cambridge, MA 02138, USA} \\
\normalsize{$^7$Institute of Astronomy, University of Cambridge, Madingley Road, Cambridge CB3 0HA, UK} \\
\normalsize{$^8$Universidad de Concepci{\'o}n, Departamento de Astronomia, Casilla 160--C, Concepci{\'o}n, Chile} \\
\normalsize{$^9$Auckland Observatory, Auckland, New Zealand} \\
\normalsize{$^{10}$Possum Observatory, Patutahi, New Zealand} \\
\normalsize{$^{11}$Farm Cove Observatory, Centre for Backyard Astrophysics, Pakuranga, Auckland, New Zealand} \\
\normalsize{$^{12}$AUT University, Auckland, New Zealand} \\
\normalsize{$^{13}$Perth Exoplanet Survey Telescope, Perth, Australia} \\
\normalsize{$^{14}$Dept.\ of Physics and Astronomy, University of Canterbury, Private Bag 4800, Christchurch, New Zealand} \\
\normalsize{$^{15}$Dept.\ of Physics and Astronomy, Texas A\&M University, College Station, Texas 77843-4242, USA} \\
\normalsize{$^{16}$Korea Astronomy and Space Science Institute, Daejeon 305-348, Republic of Korea} \\
\normalsize{$^{17}$Solar-Terrestrial Environment Laboratory, Nagoya University, Nagoya, 464-8601, Japan} \\
\normalsize{$^{18}$University of Notre Dame, Department of Physics, 225 Nieuwland Science Hall, Notre Dame, IN 46556-5670, USA} \\
\normalsize{$^{19}$Institute of Information and Mathematical Sciences, Massey University, Private Bag 102-904, North Shore Mail Centre, Auckland, New Zealand} \\
\normalsize{$^{20}$Department of Physics, University of Auckland, Private Bag 92-019, Auckland 1001, New Zealand} \\
\normalsize{$^{21}$Okayama Astrophysical Observatory, National Astronomical Observatory of Japan, Asakuchi, Okayama 719-0232, Japan} \\
\normalsize{$^{22}$Department of Earth and Space Science, Osaka University, Osaka 560-0043, Japan} \\
\normalsize{$^{23}$Nagano National College of Technology, Nagano 381-8550, Japan} \\
\normalsize{$^{24}$Tokyo Metropolitan College of Aeronautics, Tokyo 116-8523, Japan} \\
\normalsize{$^{25}$School of Chemical and Physical Sciences, Victoria University, Wellington, New Zealand} \\
\normalsize{$^{26}$Mt. John University Observatory, P.O. Box 56, Lake Tekapo 8770, New Zealand} \\
\normalsize{$^{27}$Department of Physics, Faculty of Science, Kyoto Sangyo University, 603-8555, Kyoto, Japan} \\
\normalsize{$^{28}$School of Physics and Astronomy, Tel-Aviv University, Tel-Aviv 69978, Israel} \\
\normalsize{$^{29}$Institute of Astronomy, University of Cambridge, Madingley Road, Cambridge CB3 0HA, UK} \\
\normalsize{$^{30}$Department of Earth, Ocean and Atmospheric Sciences, The University of British Columbia, Vancouver, British Columbia, V6T 1Z4, Canada}

\baselineskip24pt

\begin{sciabstract}

We detect a cold, terrestrial planet in a binary-star system using
gravitational microlensing. The planet has low mass (2 Earth masses)
and lies projected at $a_{\perp,ph}\simeq 0.8$ astronomical units (AU)
from its host star, similar to the Earth-Sun distance.  However, the
planet temperature is much lower, $T<60\,$Kelvin, because the host
star is only 0.10--0.15 solar masses and therefore more than 400 times
less luminous than the Sun. The host is itself orbiting a slightly
more massive companion with projected separation $a_{\perp,ch}=10$--$15\,\au$. 
Straightforward modification of current microlensing search strategies 
could increase their sensitivity to planets in binary systems. With more 
detections, such binary-star/planetary systems could place constraints 
on models of planet formation and evolution. This detection is consistent
with such systems being very common.

\end{sciabstract}

Although at least half of all stars are in binary or multiple systems, 
the overwhelming majority of detected exoplanets orbit single stars, or
at least stars whose companions have not been detected or are so far 
from the planet's host as to be physically irrelevant. Because binary 
stars are so common, theories of planet formation and orbital evolution 
should be strongly constrained by the observed frequency and parameter 
distributions of planets in these systems. For example, the presence 
of a relatively near companion might truncate or disrupt the proto-planetary 
disk that is thought to be the planet birthplace. Exploring planets 
in binary systems is therefore an important frontier.

Microlensing is complementary to other planet-finding techniques
in terms of sensitivity as functions of planet-host separation,
host mass, planet mass, and planet-host position within our Galaxy.
The basic scale of microlensing phenomena is set by the Einstein radius
\cite{einstein36}
\begin{equation}
\theta_\e \equiv \sqrt{\kappa M \pi_\rel};
\quad
\kappa \equiv {4 G\over c^2\au} \simeq 8.1\,{\mas\over M_\odot},
\label{eqn:thetaedef}
\end{equation}
where $M$ is the lens mass, $\pi_\rel=\au(D_L^{-1}-D_S^{-1})$ is the 
lens-source relative parallax, and $D_L$ and $D_S$ are the distances 
to the lens and source, respectively. If two stars are perfectly 
aligned on the sky, then the gravity of the one in front (``lens'') 
bends the light from the one in back (``source'') into an annulus 
(``Einstein ring'') of radius $\theta_\e$ and width twice $\theta_*$, 
where $\theta_*$ is the source angular radius. If the lens-source 
separation $\Delta\theta_{LS}$ is nonzero but still $\Delta\theta_{LS}\la\theta_\e$, 
then the source light is broken up into two images, one inside and 
the other outside the Einstein ring. The two images are separated by 
$\theta_\e \sim {\cal O}(\mas)$ ($10^{-3}$ arcseconds) and so are not 
resolvable with current telescopes. However, the combined area of 
the images is larger than the source and so appears brighter by 
the magnification $A$, which scales very nearly as 
$(\Delta\theta_{LS}/\theta_\e)^{-1}\equiv u^{-1}$ for $u\la 0.5$.
Hence, as the lens passes by the projected position of the source, 
the magnification increases and then decreases, creating a ``microlensing event''. 
Currently, over 2000 such events are discovered each year. If the lens has 
a planet, and one of the two images passes near this planet, its 
gravity further deflects the light, changing the lightcurve and 
thereby betraying its presence. Planet sensitivity peaks over the range
$0.6\theta_\e\la a_\perp/D_L\la 1.6\theta_\e$,
which corresponds to a planet-host physical separation $a_\perp$
\begin{equation}
a_\perp \sim r_\e \equiv D_L\theta_\e \sim 3.5\,\au\,\sqrt{M\over M_\odot}
\label{eqn:aperp}
\end{equation}
for typical event parameters. Because this is a 2-dimensional (2D) projection
of a 3-D elliptical orbit, with semi-major axis $a$, the semi-major axis 
is typically larger by $a/a_\perp \sim \sqrt{3/2}$, i.e. $a\sim 4.3\,\au$ 
for a solar-mass host. By contrast, the ``snow line'', outside of which 
ices can condense and so promote the growth of giant planets, is 2.7 AU 
in the solar system and is generally believed to increase monotonically 
with host mass (e.g. \cite{kk08}). Hence, microlensing probes planets 
in the cold outer regions, far from their host stars. By contrast,
the radial velocity (RV) and transit techniques are most sensitive to planets
much closer to hosts, and imaging is sensitive to planets much further out. 
Because microlensing does not depend on host (or planet) light, it is 
sensitive to low-luminosity (even non-luminous) hosts and to systems 
that are many kiloparsecs (kpc) away. Finally, in sharp contrast to all 
other methods, the amplitude of the microlensing signal does not necessarily 
decline as the planet mass decreases. This does not mean that microlensing is 
equally sensitive to all planet masses: the linear extent of the planetary caustic 
(closed curve of formally infinite magnification) declines as $\sqrt{q}$, 
where $q=m_p/M$, which reduces both the probability and duration of perturbation 
by $\sqrt{q}$. However, when these perturbations occur, they can be robustly 
detected (see review by Gaudi \cite{gaudirev}).

Microlensing event OGLE-2013-BLG-0341 was detected by the Optical Gravitational 
Lens Experiment\footnote{http://ogle.astrouw.edu.pl/ogle4/ews/ews.html/}
(OGLE, Chile) \cite{ews} and was also observed in survey mode by two other surveys, 
Microlensing Observations in Astrophysics\footnote{https://it019909.massey.ac.nz/moa/}  
(MOA, New Zealand) \cite{bond01, sumi03} and 
Wise\footnote{http://wise-obs.tau.ac.il/{$\sim$}wingspan/} \cite{sm12} (Israel). 
It was intensively observed in followup mode by six Microlensing Follow Up
Network\footnote{http://www.astronomy.ohio-state.edu/$\sim$microfun/}
($\mu$FUN) observatories. See Table 1.

\begin{table}
\begin{center}
\vskip 1em
\begin{tabular}{@{\extracolsep{0pt}}lllc}
\hline
\hline
Name        & Location   & Diam.(m) & Filter \\ 
\hline \hline
OGLE        & Chile      & 1.3      & I  \\
MOA         & New Zeal.  & 1.8      & RI \\
Wise        & Israel     & 1.0      & I  \\
Auckland    & New Zeal.  & 0.40     & R  \\
CTIO-SMARTS & Chile      & 1.3      & I  \\
Farm Cove   & New Zeal.  & 0.36     & N  \\
PEST        & West Aus.  & 0.30     & N  \\
Possum      & New Zeal.  & 0.36     & N  \\
Turitea     & New Zeal.  & 0.36     & R  \\
\hline
\hline
\end{tabular}
\end{center}
\begin{center}
Table 1. Observatories.
\end{center}
\end{table}

\begin{figure}
\centering
\includegraphics[width=15cm]{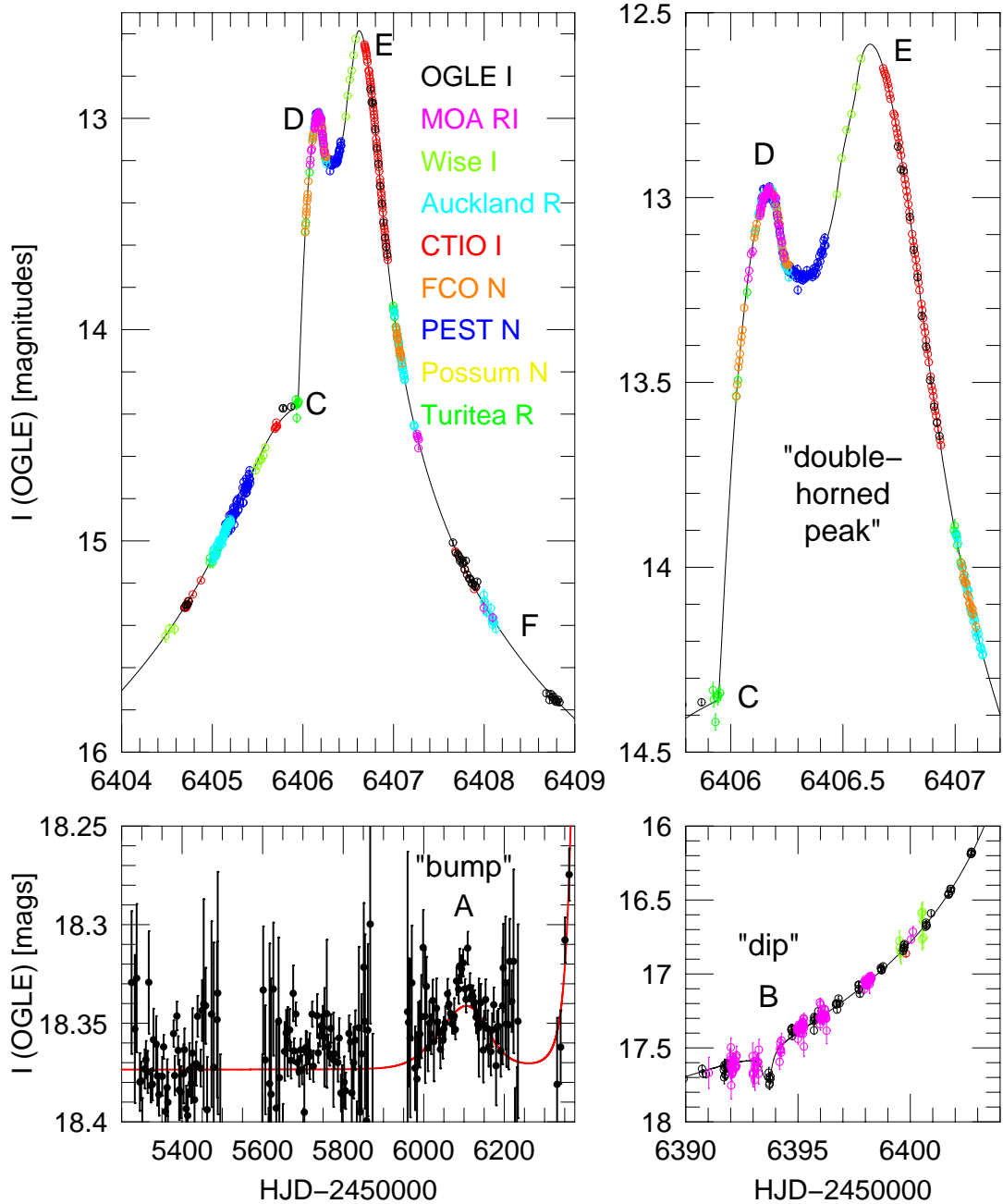}
\caption{\noindent OGLE-2013-BLG-0341 lightcurve.  Upper panels: lightcurve
features (C through F), induced by main caustic due to binary, are seen as 
source passes close to planet host. The entrance has a sharp break (C) 
indicating a caustic crossing, while the exit does not (EF), indicating a cusp exit.
Lower-left: low amplitude ``bump'' (A) due to sources passage relatively
far from binary companion to host, $\sim 300$ days earlier.  Lower-right:
``dip'' (B) due to planet ``annihilating'' one of the the two main images of the
source.
}
\label{fig:lc}
\end{figure}

\begin{figure}
\centering
\includegraphics[width=15cm]{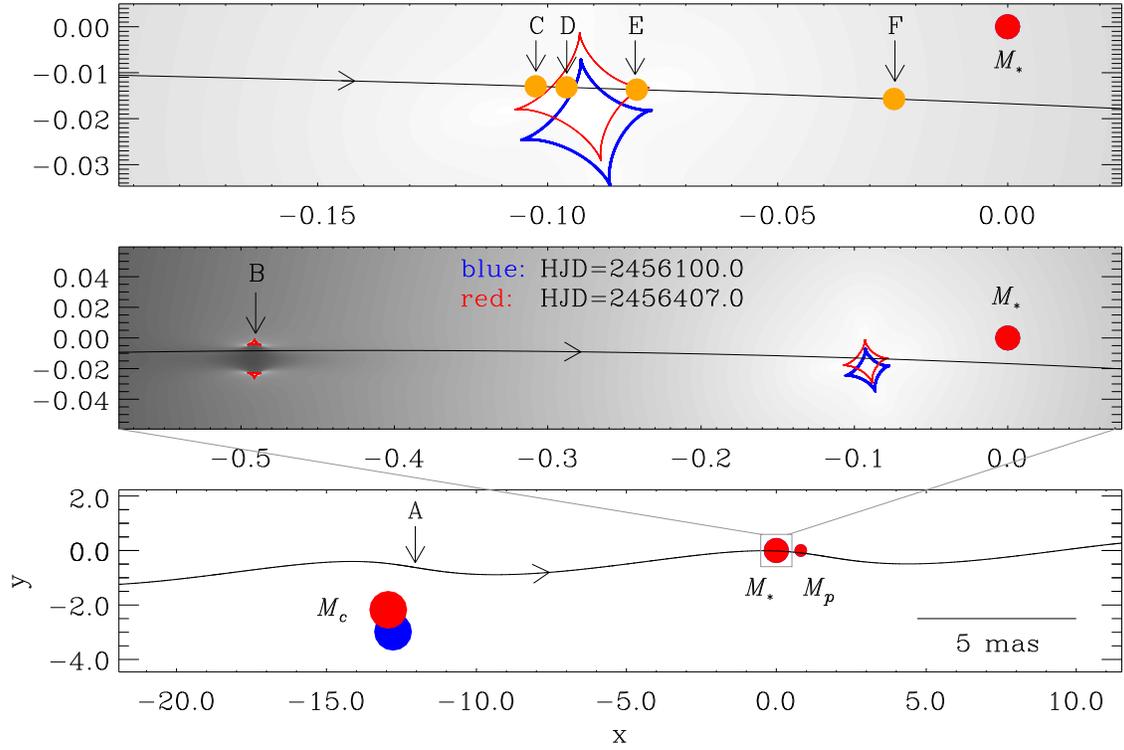}
\caption{\noindent Geometry of OGLE-2013-BLG-0341 ``Wide minus'' solution.
This includes locations of the host ($M_*$), planet ($M_p$), and companion ($M_c$),
and of the caustics (closed curves of formally infinite magnification) that 
induce strong perturbations in the lightcurve. Source position is shown at 
six key times (ABCDEF) corresponding to lightcurve features in Fig.~\ref{fig:lc}.
Middle panel: Zoom of planetary caustic (left) and central caustic (right) 
giving rise to ``dip'' and main peak seen in Fig.~\ref{fig:lc}. Central caustic 
and lens positions are shown at two different epochs (``A'' and ``E'')
separated by $\sim 300$ days during which it changed its shape and 
orientation due to binary orbital motion as described in Supplements.
Upper panel:  Further zoom showing source (yellow) to scale. Blue and red caustics 
and circles indicate lens geometries at times of ``bump'' (A) and main peak (D), 
respectively. One unit on $x$-axis corresponds to $t_\e=33\,$days in time.
}
\label{fig:geom}
\end{figure}

The precise extraction of all parameters requires computationally intensive 
modeling (see Supplements \cite{sup}), but the characteristics relevant to 
the main scientific implications of this discovery can mostly be derived from 
inspection of the light curve (Fig.~\ref{fig:lc}). There are three main 
features, a double-horned peak lasting $\sim 1\,$day centered at $t\sim 6406.5$ day, 
an extended, very shallow ``bump'' at $t\sim 6100$ day, and a very short ``dip'' 
at $t\sim 6394$ day. The first two are due to the binary and the last is 
due to the planet. Such features have been seen and analyzed previously 
in many planetary and binary microlensing events; the difference in this case 
is that 1) they appear together, and 2) there is a subtle interplay between them.  
The duration of the principal peak is $\sim 65$ times shorter than the Einstein 
diameter crossing time $2 t_\e$, where $t_\e\equiv\theta_\e/\mu_\rel \sim 33\,$days 
and $\mu_\rel$ is the lens-source relative angular speed. This peak is 
terefore due to a very small central caustic, which could in principle be due 
either to a planet near the Einstein ring, or to a companion star that is far from it.
The sharp beginning ($t\sim 6406.0$ day) and smooth end ($t>6407$ day) imply 
fold-caustic entrance and cusp exit (Fig.~\ref{fig:geom}). This morphology is 
consistent only with a binary lens. Although the binary companion could be, 
in theory, either very far inside (``close'') or very far outside (``wide'')
the Einstein ring, the early bump at $t\sim 6100$ day confirms the latter 
interpretation: this bump was generated by the source passing moderately 
close to the companion a year earlier (Fig.~\ref{fig:geom}).  Finally, 
the small dip at $t\sim 6394$ day can only be caused by a planet that is 
inside the Einstein ring. Recall that the principal lens creates two images, 
which are at extrema of the time-delay surface (Fermat's Principle). 
The outside image is at a minimum of this surface, and the inside image
is at a saddle point. A planet sitting at exactly this saddle point will 
effectively annihilate the image, causing a dip. To generate only a dip and 
no neighboring bumps, the source must have ``threaded'' the planetary-caustic 
structure as it headed toward the central caustic (Fig.~\ref{fig:geom}).  
The half-crossing time of the dip is $t_{\rm dip}\sim 0.25\,$day. Because 
the planetary caustic size scales as $(t_{\rm dip}/t_\e)\sim q^{1/2}$, 
the planet is $q\sim 0.6\times 10^{-4}$ times less massive than its host. 
From the fact that the interval between the planetary and binary caustics is 
$\sim 0.4\,t_\e$, we calculate that the planet-host separation (normalized 
to the Einstein radius) is $s_2\sim 1 -0.4/2=0.8$ from the center of magnification 
of the system (which in this binary system is very close to the host).

The next step is to transform the dimensionless separations into angles 
by measuring $\theta_\e$, using the source size $\theta_*$ as a ``ruler'' 
\cite{ob03262}. From its measured color (and thus, since stars are 
approximate black bodies, surface brightness, $S$) and flux $F$, 
we determine $\theta_* =\sqrt{F/\pi S}= 2.9~\mu{\rm as}$ \cite{sup}.
Comparing the caustic rise time (6405.97--6406.17) to the standard
mathematical form (steep, then rounded rise totaling 1.7 source-radius 
crossing times, \cite{gouldandronov}), and for simplicity ignoring that 
this entrance is at an angle, we can estimate a source crossing time of
$t_*\sim 0.12\,$days. The resulting Einstein radius is 
$\theta_\e=(t_\e/t_*)\theta_*\sim 0.8\,\mas$.

\begin{figure}
\centering
\includegraphics[width=15cm]{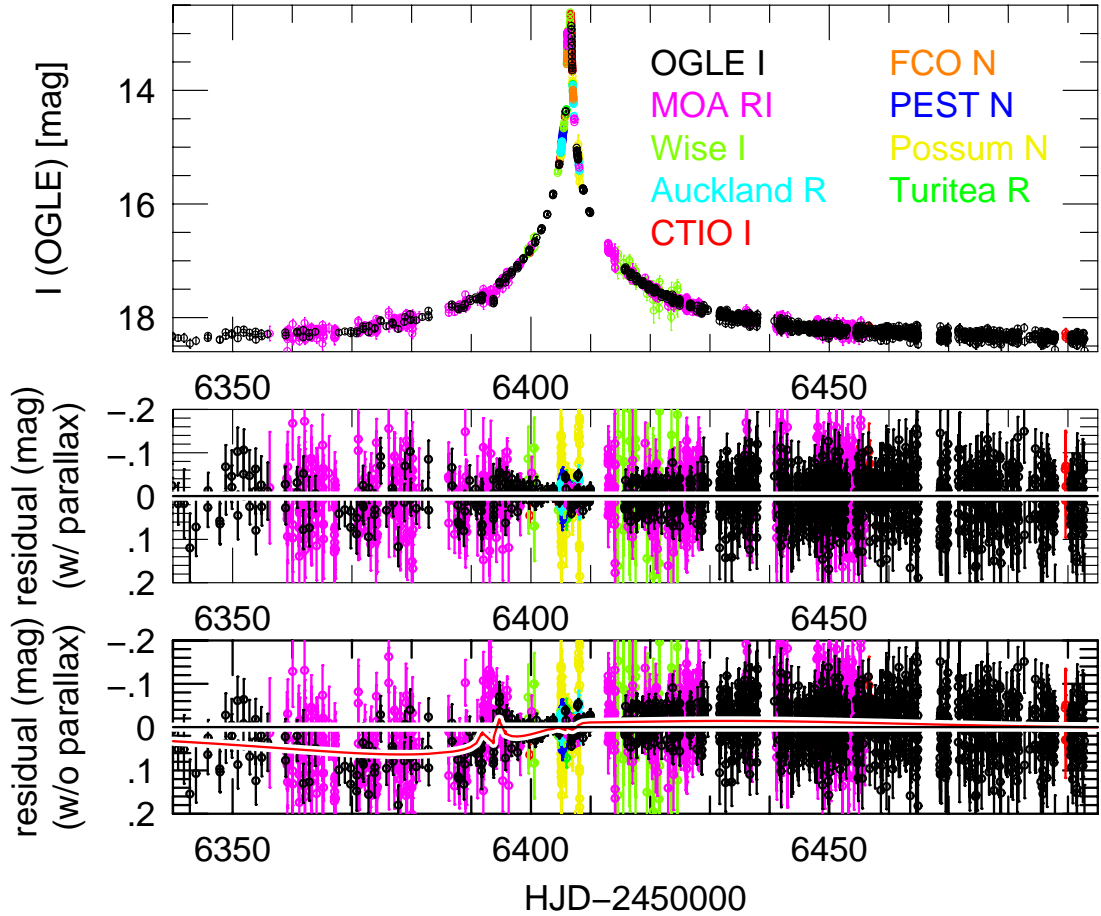}
\caption{\noindent Full OGLE-2013-BLG-0341 lightcurve (top) and residuals from 
``Wide minus'' model with (middle) and without (bottom) including the parallax effect. 
Parallax is strongly detected, $\Delta\chi^2=730$. Silhouetted black and red curves 
indicate zero and difference between parallax and no-parallax models, respectively.
In contrast to all other crucial lightcurve parameters, the parallax effect is 
not directly visible in the lightcurve, but only in the residuals. 
However, as explained in Supplements, an experienced modeler can ``read off'' 
from these residuals that $\pi_\e \ga 0.7$.
}
\label{fig:lcpar}
\end{figure}

The final step is to measure the distance using the ``microlens parallax'',
$\pi_\e\equiv \pi_\rel/\theta_\e$. This quantifies the amplitude of
lens-source relative motion due to reflex motion of Earth's orbit (scaled 
to the Einstein radius) and therefore the amplitude of the lightcurve deviations
due to this effect (see \cite{gouldhorne}, Fig.~1 for a didactic explanation). 
The impact of this effect on the lightcurve is easily seen (if not quantified) 
in the residuals to models with and without parallax (see Fig.~3). There is a 
well known degeneracy in parallax solutions (labeled ``+/-''), depending on 
which side of the projected position of Earth the lens passes relative to 
the source \cite{smith03}, and we show \cite{sup} that this degeneracy cannot be 
broken in this case\footnote{We also include in the model orbital motion of 
the binary about its center of mass. These effects are discussed in detail 
in the Supplement, but are too subtle to detect by eye from the lightcurve.}.  
Thus, we find $\pi_\e=1.0$ or $\pi_\e=0.8$. Then from the definitions of 
$\theta_\e$ and $\pi_\e$, we obtain $\pi_\rel=\theta_\e\pi_\e$ and 
$M=\theta_\e/\kappa\pi_\e$ and have calculated the relevant physical parameters 
that are derived from each of the two solutions (Table~2).

\begin{table}                                                                  
\begin{center}
\vskip 1em                                                                     
\begin{tabular}{@{\extracolsep{0pt}}llcc}                                      
\hline                                                                         
\hline                                                                         
Parameter & Unit & Wide(+) & Wide(-) \\                                        
\hline \hline                                                                  
$M_{\rm total}$           & $M_\odot$  & 0.2336 & 0.3207 \\
                          &            & 0.0182 & 0.0284 \\
\hline                                                   
$M_{\rm host}$            & $M_\odot$  & 0.1127 & 0.1452 \\
                          &            & 0.0089 & 0.0135 \\
\hline                                                 
$M_{\rm planet}$          & $M_\oplus$ &   1.66 &   2.32 \\
                          &            &   0.18 &   0.27 \\
\hline                                                 
$M_{\rm companion}$       & $M_\odot$  & 0.1209 & 0.1755 \\
                          &            & 0.0094 & 0.0155 \\
\hline                                                 
Distance                  & kpc        &  0.911 &  1.161 \\
                          &            &  0.070 &  0.093 \\
\hline                                                 
$a_{\perp, ph}$           & AU         &  0.702 &  0.883 \\
                          &            &  0.022 &  0.043 \\
\hline                                                 
$a_{\perp, ch}$           & AU         & 10.536 & 14.013 \\
                          &            &  0.318 &  0.617 \\
\hline
\hline
\end{tabular}\par
\end{center}
Table 2. Physical Parameters of the binary + planet models.
Masses, Distances, and projected separations for binary+planet system 
in the two models that are consistent with the microlensing data.
\end{table}

In either case, the planet has mass $m_p\sim 2$ Earth masses ($M_\oplus$) and 
the host is a late M dwarf, with another, slightly more massive, M dwarf as 
a companion lying at a projected separation of 10 or 14 AU. The entire system 
lies $\sim 1\,$kpc from the Sun. Simulations of microlensing with realistic 
planetary systems that include eccentricity and inclination \cite{zhu14}
confirm the naive expectation that these projected separations are good proxies 
for the semi-major axis (after upward adjustment by $\sqrt{3/2}$ to correct 
for projection effects). Indeed, the relation between $a$ and $a_\perp$ is 
very similar to the relation between $m$ and $m\sin i$ for RV detections. 
Hence, as with RV masses, this proxy can fail badly in rare individual cases.

What are the implications? First, while we cannot reliably estimate the frequency 
of such systems, we can ask the simpler question: if all stars were in such 
binary/terrestrial-planet systems, how many should have been detected? 
The detection required 1) a transit of the source by both the planetary 
$(p\sim 6\times 10^{-3})$ and central caustics $(p\sim 7\times 10^{-2})$ and 
2) relatively high-cadence data on a relatively bright star $I_S<18.5$ to ensure
sufficient signal-to-noise ratio to detect the dip. Therefore, if all $I<18.5$ 
stars that undergo microlensing events had such planets, we would detect 
$\sim 0.04\%$ of them. During the three years that OGLE-IV has issued alerts, 
it detected a total of $\sim 10^3$ of $I_S<18.5$ events in its high-cadence fields, 
which implies an expectation of 0.4 such planets. This would be compatible with 
survey results showing that Earths and super-Earths are the most common type of 
planet orbiting stars with a wide range of masses 
\cite{mayor11,dong13,petigura13,swift13,dressing13} 
and with predictions from microlensing based on more massive planets orbiting 
low-mass stars \cite{cassan12}.

Second, this result shows that terrestrial planets can exist relatively far 
($\sim 1\,\au$) from their hosts even if the latter have relatively nearby 
($\la 20\,\au$) binary companions, thus providing empirical test of models of 
terrestrial planet formation in such close binaries (e.g., \cite{barbieri02,quintana07,rafikov13}).

Third, when combined with the RV detection of a terrestrial ($m_p\sin i=1.3\,M_\oplus$) 
planet orbiting very close (0.04 AU) to $\alpha$CenB \cite{alphacenb}, which is 
a solar-type star, it shows that terrestrial planets can form in binaries with
diverse properties in terms of host mass and planet-host separation. Although 
OGLE-2013-BLG-0341LBb was discovered in a search of $\sim 10^3$ microlensing 
events and $\alpha$CenBb resulted from intensive observations of a single system, 
the expected yield in each case (if all stars had similar planets) was roughly unity.

Planets have been discovered in a variety of binary configurations. For example, 
about 7 transiting circumbinary planets have been discovered in {\it Kepler}
satellite data \cite{welsh12}, and two Jovian planets have been found in binary 
systems using RV \cite{gammacepb,hd41004}. Microlensing is also sensitive to 
planets in very different binary configurations, and both current and future 
surveys are likely to discover these.

Finally, we discuss an extremely interesting aspect of the modeling of 
OGLE-2013-BLG-0341 that points to the possibility of much greater sensitivity 
to systems of this type. When the data near the dip are removed and 
the remaining lightcurve is fitted for a binary both with and without a planet, 
the former solution is preferred by $\Delta\chi^2=216$ over the no-planet model. 
That is, although the planet is lighter than the binary by a factor $q<10^{-4}$, 
its presence distorts the caustic enough to be noticed in the very high-density 
observations of the caustic features. Moreover, this model accurately 
``predicts'' the position of the planet. This means that the planet 
could have been detected even if the source had not passed over the tiny 
planetary caustic. This passage accounted for $p\sim 1/170$ in the above
probability calculation, implying that by probing the central caustic,
sensitivity can be improved by $1/p\sim 170$. However, high-density
observations (as in Fig.~\ref{fig:lc}) are not routinely taken for binaries. 
Indeed, $\mu$FUN organized these only because it recognized from the form of 
the planetary caustic that the source was headed toward the central-caustic 
region and sought to exploit this passage to obtain information about the planet.
The resultant dense coverage of the binary caustic was inadvertent.
Dense followup of ``ordinary'' binaries may then be the best way to probe 
for planets in binary systems \cite{lee08}. Because there are a comparable 
number of high-magnification binary compared to apparently-single-star events, 
the additional observing resources required to carry out such followup is 
relatively modest.

\clearpage
\bigskip
\noindent
{\bf Supplementary Materials}\\
{\tt www.sciencemag.org}\\
Materials and Methods \\
Supplementary text\\
Figs. S1, S2, S3\\
Table S1\\
References ($28$--$47$)

\clearpage

\begin{center}
\title{{\huge Supplementary Materials for}\\[0.5cm]
{\bf\large A Terrestrial Planet in a $\sim 1\,\au$ Orbit
Around One Member of a $\sim 15\,\au$ Binary}} 
\end{center}

\section{Data Collection and Initial Reductions}

Data for OGLE-2013-BLG-0341 (RA = 17:52:07.49, Dec = $-29$:50:46.0)
$(l,b)=$ (-0.05,-1.68)
were obtained through a complex interplay
of three modes, real time alerts of ongoing microlensing events,
high-cadence monitoring of high magnification events, and intensive 
monitoring of wide microlensing fields, which appeared historically
in that order.  The \cal{O}$(1\,{\rm day}^{-1})$ cadences of early surveys
were too low to robustly detect and characterize the $\sim 1\,$day 
planetary perturbations, but since the survey teams using wide-angle
cameras were able to detect ongoing events in real time \cite{ews}, 
follow-up teams formed to intensively monitor a subset of events using
networks of narrow-angle cameras, as suggested by \cite{gouldloeb92}.
When the OGLE survey upgraded from OGLE-II to OGLE-III, thereby increasing
event detection by a factor $\sim 10$, it became practical for follow-up
teams to focus on rare high-magnification events, which are substantially
more sensitive to planets \cite{griest98}.  This became the main
channel of microlensing planet detection during 2005-2010.  See \cite{gould10}
for a thorough review.  Further upgrades to OGLE-IV and MOA-II, and
the addition of the Wise survey, which all employ very large format cameras, 
enabled moderately high-cadence ($\ga 1\,{\rm hr}^{-1}$) near round-the-clock
monitoring of $\ga 10\,{\rm deg}^2$ of the densest star fields.  This
permits planet detection in a much larger number of events without any
conscious human intervention (e.g., \cite{ob120406}) thus both greatly
increasing the planet detection rate and permitting more rigorous statistical
analysis.  The introduction of such ``second generation'' surveys was
once thought to eliminate the need for follow-up observations.

Indeed, the binary+planet nature of OGLE-2013-BLG-0341 could have been
established based on survey data alone.  Data for the both the ``bump''
(A) and the ``dip'' (B) were collected without conscious intervention,
and survey data by themselves would have covered the
central binary caustic (C--F) well enough to basically characterize it.
Nevertheless, by analyzing ongoing OGLE data, which were posted to the
web daily, $\mu$FUN detected subtle signs of an approaching caustic
and organized intensive observations to capture the entrance (if it
occurred).  Coordinated efforts of several amateur astronomers in New Zealand,
some looking for holes in clouds through which to observe the event,
captured this entrance, thereby pinning down the caustic structure
much more precisely.  This detection then further triggered extremely
dense observations, up to $\sim 20\,{\rm hr}^{-1}$.  It is this
extremely dense coverage that permits measurement of such detailed
effects as orbital motion and the presence of the planet from 
central-caustic data alone.  See below.  In fact, there are several
recent events that demonstrate synergy between second-generation surveys
and the high-magnification follow-up observations that were previously
thought to be a relic of first-generation microlensing
\cite{mb11293,ob120026,mb13220}.

The data were reduced using ``image subtraction'' (aka ``difference
image analysis'') in which successive images are geometrically and
photometrically aligned, convolved to a common point spread function
(PSF) and then subtracted from a reference image \cite{alardlupton}.  
In principle (and
very nearly in practice), this leads to a completely flat ``difference
image'' except where some source has varied, either by changing
brightness or by moving between images.  The major exception is residuals
from bright stars due to imperfect modeling of the PSF.  This
technique is very important for microlensing observations, which
take place in the densest star fields on the sky where it is actually
quite rare for a microlensed source not to be blended with a
random star along the line of sight.  Difference imaging ``magically''
removes essentially all such irrelevant blends (but see below).

\section{Initial Modeling and Re-reductions}

Even as OGLE-2013-BLG-0341 was still falling toward baseline,
initial models were circulated with the same overall characteristics
and similar model parameters to those reported here.  It was noted
at that time that both close-binary and wide-binary models gave
excellent fits to the 2013 data.  However, it was then noticed that
the wide models ``predicted'' an earlier passage near the putative
wide companion roughly one year earlier.  Binning these data revealed
a low-amplitude but highly significant ``bump'' as seen in Fig.~1.

Nevertheless, the early (pre-event) data showed a number of puzzling
features that might call into question the reality of this ``bump''
(and so of the wide-binary interpretation).  The OGLE data, which
by chance extend back about 15 years into both OGLE-III and OGLE-II,
show a steady brightening of about 0.4\% per year.  Of course, this
effect is too small to be seen in individual data points but is plainly
visible in binned data.  This raised two concerns.  First, if the
source were variable on long timescales, then it might also have
varied on shorter timescales one year before the event, thus giving
rise to the ``bump''.  Second, whatever the origin of this variation, it
could affect the estimation of ``baseline flux'', which itself is not
of any direct interest but can impact other event parameters, which
are of interest (e.g., \cite{mb11293}).  Hence, a decision was made
to wait until the event had effectively returned to baseline to complete
the analysis and, in the meantime, to try to track down the origin of the
observed long-term variability.  
An important clue in this regard is that MOA data
showed a similar long-term trend and also showed
strong variation as a function of position of the source relative to
the ground (which changes as functions of time of night and time of year).

We began by consulting variable-star experts who told us flatly that
it was extremely unlikely for this type of star (reddish subgiant) to
be varying on long timescales at few percent levels.  

An intensive investigation revealed that a neighboring star 
(about 1.5 times the source brightness and separated from it
by $1.3^{\prime\prime}$) was slowly moving toward the microlensed source
at $5\,\masyr$.  In the difference images, this star then produced
an extremely small dipole ``divot'', with an excess flux near the 
microlensed source and an exactly equal deficit further from it.
The excess then entered the tapered aperture used to measure the
microlensed source but the deficit did not.  As the neighboring star
moved closer, the microlensed source appeared to brighten.  Detailed
modeled showed that this effect completely explained the apparently 
increasing brightness.

The MOA data have substantially larger PSF 
than OGLE (due to much better observing conditions in Chile than NZ), 
meaning that the two sides of the ``divot'' move in and out
of tapered aperture as the PSF varies.  This explains the much
stronger PSF-effects seen in MOA data, while still accounting for the
long term trend.  In addition, the larger PSF, together with differential
refraction, explains the variations as a function of position relative
to the ground.  This is seen in many other events and is due simply to
the presences of the neighbor, not its motion.

The bottom line is that these effects can be robustly removed from the
OGLE data, and can be accounted for in the MOA data.  We therefore
adjusted the OGLE data to remove this trend, but only used the MOA
data within 50 days of the peak, in order to guard against unmodeled
effects of neighbor motion in the MOA baseline.  Note that we did not
check the impact of excluding MOA baseline data, but made the decision
solely on the grounds that the corrections were substantially less reliable
than for OGLE data.

Rereduced data are available from the corresponding author at the following site. \\
{\tt http://astroph.chungbuk.ac.kr/$\sim$cheongho/OB130341/data.html}

\section{Final Modeling}

\subsection{Event Parameters}

Fourteen geometric parameters are required to model the OGLE-2013-BLG-0341
lightcurve, in addition to two flux parameters for each observatory.
That is, the observed flux $F_i$ is modeled as 
$F_i(t)= F_{s,i}A(t) + F_{b,i}$ where $(F_s,F_b)_i$ are the source and
blended fluxes at observatory $i$ and $A(t)$ is the (very nearly, see below)
observatory-independent magnification.

The overall microlensing event is characterized by 
the Einstein radius crossing time,
$t_\e$, the impact parameter $u_0$ of the source relative to the host 
star (designated ``primary'') in units of $\theta_\e$, 
and the time $t_0$
of this approach.  The primary's two companions
are each described by their mass ratios $(q_2,q_3)$ and 
projected separations $(s_2,s_3)$ in units of $\theta_\e$ 
relative to the
primary.  The angle between them is $\phi_{23}$,
while the angle between the source trajectory and the primary-planet axis
is $\alpha$.  Next,
$\rho \equiv \theta_*/\theta_\e$ where $\theta_*$ is the
angular source radius.  Equivalently, $t_*\equiv \rho t_\e$ is the
source-radius self-crossing time.

The reflex motion of Earth's orbit affects the lens-source separation
enough to measure the two ``microlens parallax''
parameters $\bpi_\e=(\pi_{\e,N},\pi_{\e,E})$ 
\cite{gould92,gould04},
whose magnitude is $\pi_\e = \pi_\rel/\theta_\e$, and whose
direction is that of the lens-source relative proper motion:
$\bpi_\e/\pi_\e = \bmu/\mu$.   Finally, in many binary and planetary
events, it is possible to detect the instantaneous projected orbital
motion of the lens components
$\bgamma=(\gamma_\parallel,\gamma_\perp)=((ds/dt)/s,d\psi/dt)$,
where $\psi$ is the angular orientation of the binary axis.  In
hierarchical triples, there could be two such $\bgamma$ (although
we hold the planet position fixed relative
to the host, allowing only binary-star orbital motion).

The microlens parallax and orbital motion is subject to another 
degeneracy that takes lensing parameters
$(u_0,\alpha,\pi_{\e,\perp},\gamma_\perp)$ $\rightarrow$
-$(u_0,\alpha,\pi_{\e,\perp},\gamma_\perp)$,
where
$\pi_{\e,\perp}$ is the component of $\bpi_\e$ perpendicular to the
projected position of the Sun \cite{smith03,skowron11}.  See \cite{skowron11}
for a thorough review of this parameterization.

\subsection{Modeling Methods}

The modeling of the lightcurve is complex and computationally intensive for
two interrelated reasons.  First, while the majority of the individual
data points lie far from the caustics (and so can be evaluated 
solving a 10th order complex polynomial) or are moderately near
(and so solvable using 13 such evaluations in the hexadecapole
approximation \cite{pejcha09,gould08}), there are a very large number
of points within or very near the caustics, which must be evaluated
using ``inverse ray shooting'' \cite{schneider86,schneider87}.
In this approach, one ``shoots'' rays back from the observer and calculates
where they land on the source plane due to deflections induced by the three
bodies.  Those landing on the source are weighted by the local
surface brightness, which is affected by limb darkening.  We use the
linear limb-darkening coefficients $u=(0.649,0.702,0.754,0.754)$
\cite{claret00}
for $(I,R/I,R,N)$ filters based on stellar parameters 
$(T_{\rm eff},\log g)=(4250,3.00)$ found from the source position
on the color-magnitude diagram (below), and adopt turbulence of $2\,\kms$
and solar heavy element abundance (neither of which have any significant
effect).

Second, the large number of parameters, together with the presence 
of sharp features in the lightcurve, can cause ``downhill'' algorithms
to become stuck in false minima.  

We therefore organized two completely independent searches of this
parameter space by two teams within our collaboration, 
using two completely independent numerical algorithms.  Another
collaboration member, not involved in either calculation, then
collected the results and led in the resolution of the modest differences.

\begin{figure}
\begin{center}
\includegraphics[width=16cm]{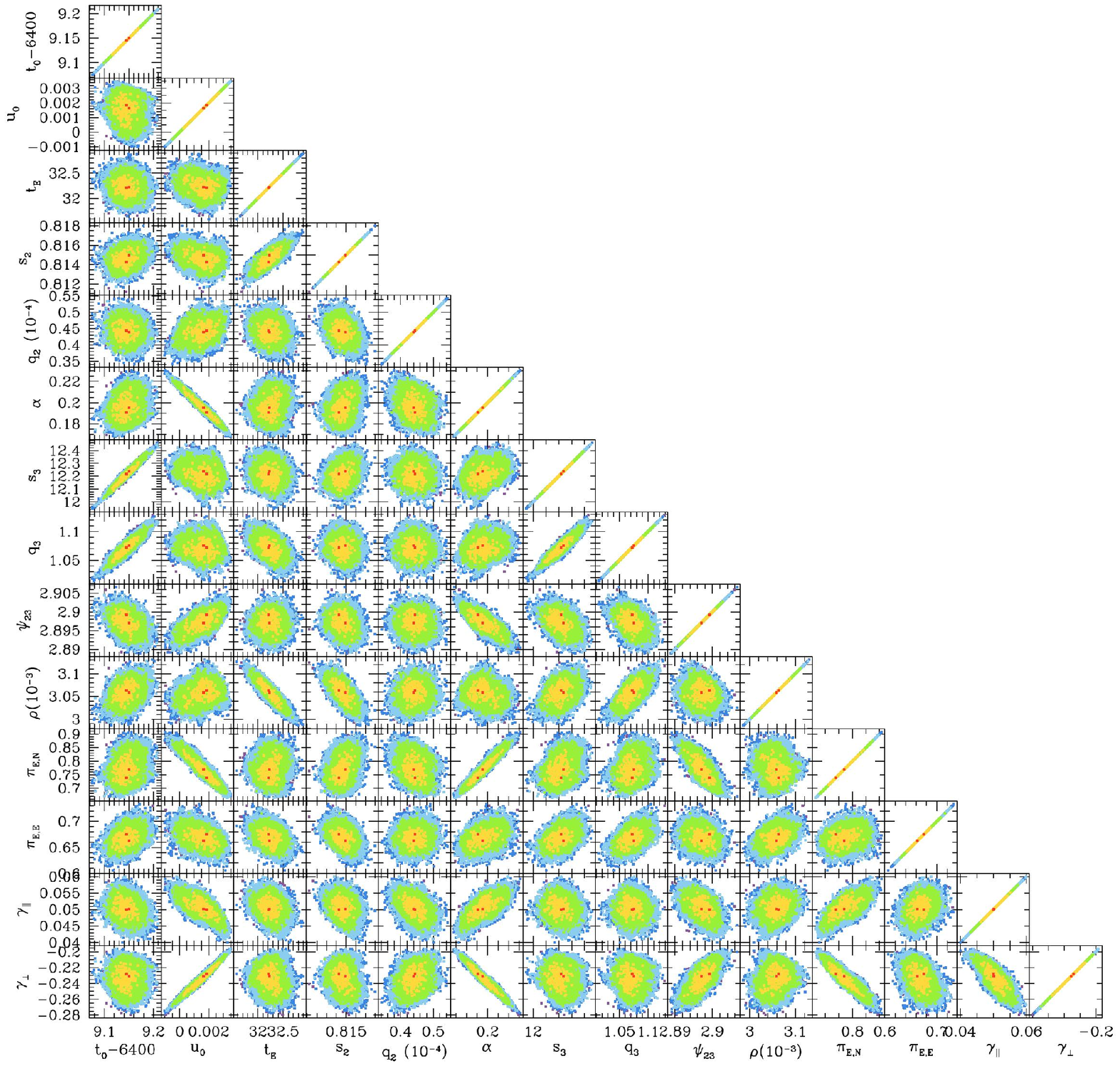}
\end{center}
Figure S1: 
Posterior distribution of 14 microlensing parameters of ``wide (plus)''
solution, whose central
values and errors are shown in Table S1.  Color coding indicates
points on the Markov Chain within 1 (red), 2 (yellow), 3 (green), 
4 (cyan), 5 (blue) sigma of the best fit.
\label{fig:w+}
\end{figure} 

\begin{figure}
\begin{center}
\includegraphics[width=16cm]{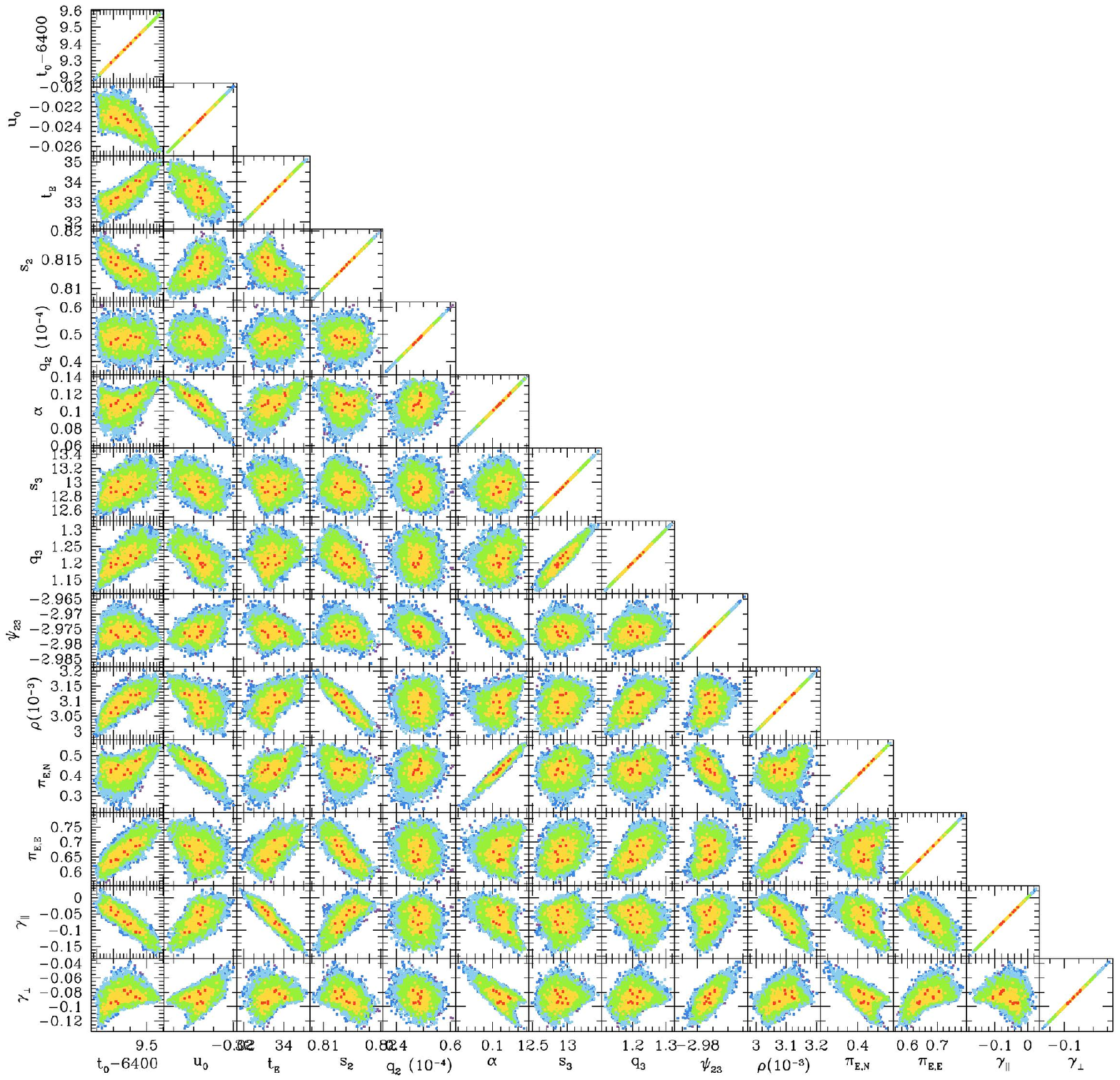}
\end{center}
Figure S2: Same as Fig.~S1, but for ``wide (minus)'' solution.
\label{fig:w-}
\end{figure}

\begin{table}                                                                  
\begin{center}
\vskip 1em                                                                     
\begin{tabular}{@{\extracolsep{0pt}}llrrrrr}                                   
\hline                                                                         
\hline                                                                         
Parameter & Unit & Close(+) & Close(-) & Wide(+) & Wide(-) & Wide*(-) \\       
\hline \hline                                                                                         
$\chi^2/$dof        &                     &    9108 &    9086 &    8876 &    8900 &    8749 \\
                    &                     &  / 8889 &  / 8889 &  / 8889 &  / 8889 &  / 8755 \\
\hline                                                                                        
$t_0 - 6400$        &day                  &   2.280 &   2.239 &   9.148 &   9.362 &   9.274 \\
                    &                     &   0.049 &   0.048 &   0.018 &   0.078 &   0.021 \\
\hline                                                                                        
$u_0$               &                     &  0.0347 & -0.0334 &  0.0013 & -0.0233 & -0.0238 \\
                    &                     &  0.0005 &  0.0004 &  0.0008 &  0.0010 &  0.0005 \\
\hline                                                                                        
$t_{\rm E}$         &day                  &   31.40 &   31.97 &   32.26 &   33.41 &   32.74 \\
                    &                     &    0.23 &    0.23 &    0.16 &    0.61 &    0.08 \\
\hline                                                                                        
$s_2$               &                     &   0.949 &   0.951 &   0.815 &   0.814 &   0.808 \\
                    &                     &   0.002 &   0.002 &   0.003 &   0.007 &   0.008 \\
\hline                                                                                        
$q_2$               &$10^{-4}$            &   1.806 &   1.681 &   0.443 &   0.480 &   0.468 \\
                    &                     &   0.157 &   0.154 &   0.029 &   0.033 &   0.012 \\
\hline                                                                                        
$\alpha$            &radian               & -0.0026 & -0.0065 &  0.1990 &  0.1068 &  0.0871 \\
                    &                     &  0.0028 &  0.0020 &  0.0106 &  0.0107 &  0.0073 \\
\hline                                                                                        
$s_3$               &                     &  0.1880 &  0.1863 & 12.2274 & 12.9186 & 13.0299 \\
                    &                     &  0.0008 &  0.0009 &  0.0334 &  0.0641 &  0.0287 \\
\hline                                                                                        
$q_3$               &                     &   2.429 &   2.420 &   1.073 &   1.211 &   1.215 \\
                    &                     &   0.054 &   0.054 &   0.015 &   0.031 &   0.026 \\
\hline                                                                                        
$\phi_{23}$         &radian               &  0.2765 & -0.2598 &  2.8967 & -2.9756 & -2.9503 \\
                    &                     &  0.0080 &  0.0065 &  0.0024 &  0.0025 &  0.0089 \\
\hline                                                                                        
$\rho$              &$10^{-3}$            &  3.1779 &  3.1032 &  3.0543 &  3.0943 &  3.0466 \\
                    &                     &  0.0274 &  0.0268 &  0.0159 &  0.0276 &  0.0134 \\
\hline                                                                                        
$\pi_{\rm E,N}$     &                     &  -1.740 &   1.557 &   0.785 &   0.421 &   0.440 \\
                    &                     &   0.160 &   0.141 &   0.046 &   0.048 &   0.015 \\
\hline                                                                                        
$\pi_{\rm E,E}$     &                     &   0.692 &   0.981 &   0.671 &   0.671 &   0.631 \\
                    &                     &   0.037 &   0.053 &   0.016 &   0.038 &   0.010 \\
\hline                                                                                        
$\gamma_\parallel$  &${\rm yr}^{-1}$      &  2.5944 &  0.7552 &  0.0507 & -0.0650 & -0.0200 \\
                    &                     &  0.9863 &  1.1885 &  0.0030 &  0.0346 &  0.0014 \\
\hline                                                                                        
$\gamma_\perp$      &${\rm yr}^{-1}$      &  1.1523 & -0.1799 & -0.2367 & -0.0852 & -0.1065 \\
                    &                     &  0.2201 &  0.1773 &  0.0150 &  0.0118 &  0.0050 \\
\hline                                                                                        
$I_s$               &                     & 18.5881 & 18.6103 & 18.6130 & 18.6028 & 18.6190 \\
                    &                     &  0.0088 &  0.0088 &  0.0066 &  0.0111 &  0.0081 \\
\hline                                                                                           
$\beta=(E_{\rm kin}/E_{\rm pot})_\perp$&  & 0.00162 & 0.00033 & 3.70949 & 1.34161 & 1.35543 \\
                    &                     & 0.00105 & 0.00048 & 0.34936 & 0.41431 & 0.15313 \\
\hline
\hline
\end{tabular}\par 
\end{center}
\smallskip
Table S1. OGLE-2013-BLG-0341 microlensing parameters.
Best fit values and $1\,\sigma$ error bars for the 14 microlensing model parameters 
described in Section 3.1, the OGLE source magnitude, and the energy 
parameter $\beta$ described in Section 5.2.  The first two model columns 
are ``close'' binary solutions (both bodies inside the Einstein radius), while the 
next two show ``wide'' solutions (one body outside).  The ``$(+/-)$'' solutions refer 
to the lens passing on different sides of Earth.  The last column shows the best 
``wide(-)'' solution with 144 points near the ``dip'' removed from the data.  
It is nearly identical to the ``wide(-)'' solution with all data.
\end{table}

We label the four solutions Wide (+/-), and Close (+/-).  In the wide
solutions, the binary companion lies well separated from the host while
in the close solution it lies well inside the projected position of the
planet.  The best fit parameters for each of these four solutions
are shown in Table S1.  In addition, we show a fifth solution
``wide*(-)'', which has the ``wide(-)'' geometry but for which 
144 points in and near
the ``dip'' have been removed from the data.  Note that the 3 planet parameters
are nearly identical for this solution as the regular ``wide(-)'' solution.
$(s_2,q_2,\phi_{23})$.  This implies that the planet could have been
detected and characterized even if the source had missed the planetary
caustic (so, no ``dip'').  
Fig.~S1 and S2 each show $14\times 13/2=91$ 2-dimensional
slices through the posterior distributions of the wide(+) and wide(-)
solutions, respectively.  
Note that these are each well localized relative to the
parameter values, except for $\gamma_\parallel$, whose main interest is that
it is near zero (see
below).  Because of this compactness, the choice of priors plays very
little role.  We used flat priors.

\section{Source Characteristics}

\begin{figure}
\begin{center}
\includegraphics[width=15cm]{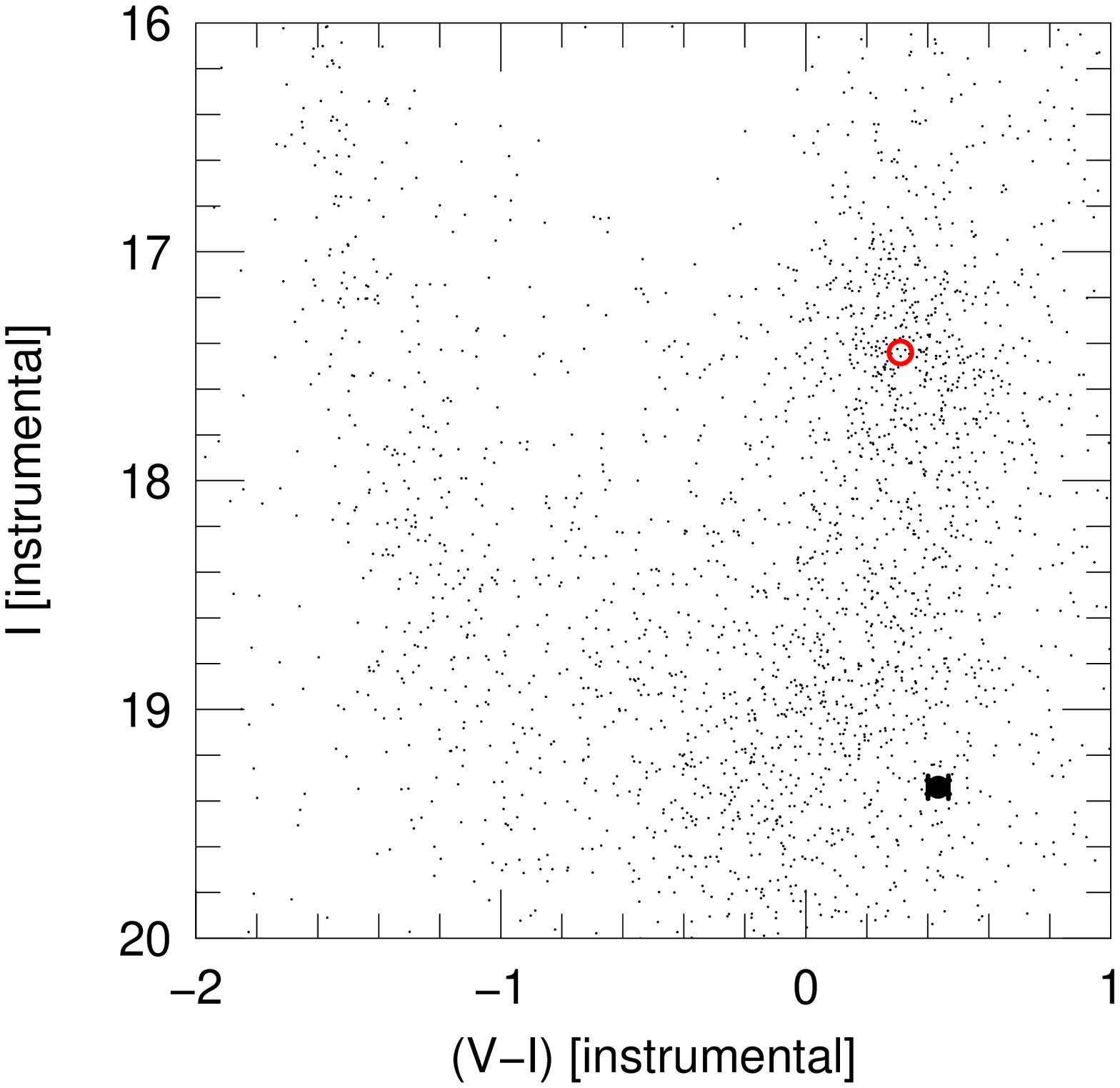}
\end{center}
Figure S3: Instrumental CTIO color-magnitude diagram (CMD)
showing the position of the OGLE-2013-BLG-0341 baseline source (black)
compared to all stars within $90^{\prime\prime}$.  The CMD is deliberately
not calibrated because all that is of interest is the difference
between the source (black) and the centroid of the red clump (red):
$\Delta(V-I,I) = (0.12,1.90)$.  Together with the known position
of the clump $(V-I,I)_{0,\rm cl} = (1.06,14.45)$,
this implies $(V-I,I)_{0,s} = (1.18,16.35)$,  Combined with a similar
measurement from OGLE and the color/surface-brightness relation 
\cite{kervella04}, this position implies an angular source radius 
$\theta_*=2.89\pm 0.23\,\muas$.
\label{fig:cmd}
\end{figure} 

In order to measure the Einstein radius $\theta_\e=\theta_*/\rho$,
one must determine the source radius $\theta_*$.  We follow the
standard procedure \cite{ob03262}.  First, we measure the source
flux in $V$ (not used in modeling but just for this purpose) and $I$,
which come directly out of the modeling (see above).  We then
plot the resulting $(V-I,I)_s$ on an instrumental color-magnitude
diagram of the field and measure the offset from the
centroid of the red giant ``clump''. See Fig.~S3.  If this
diagram were calibrated, then the positions of the source and
clump would both change, but the offset would not.  Next, we use
the dereddened color \cite{bensby13} and magnitude \cite{nataf13}
of the clump $(V-I,I)_{0,\rm lc}=(1.06,14.45)$
together with this offset to find the dereddened source $(V-I,I)_{s,0}$.
This procedure is known to reproduce the true dereddened color to $\pm 0.05$
\cite{bensby13}.  We estimate an error in $I_{s,0}$ of $\pm 0.10$ magnitudes
based on reproducibility of centroiding the clump.  We convert from
$V/I$ to $V/K$ using the color-color relations of \cite{bb88}.  Finally,
we use the empirically determined color/surface-brightness relations of
\cite{kervella04} to determine $\theta_*$.  We obtain $(V-I)_0=1.18\pm 0.01$
from CTIO data and $(V-I)_0=1.15\pm 0.03$ from OGLE data, and $I_0=16.35$,
which (after taking account of the above 0.05 and 0.10 errors) yields
$\theta_*=2.89\pm 0.23\,\muas$.

\section{Higher-Order Effects}

There are two higher-order effects that are not obvious from the
lightcurve, but have important physical implications: parallax
and binary orbital motion.  

\subsection{Parallax}

As illustrated in Fig.~3, if parallax is not included in the model,
the residuals are severe:  including parallax reduces $\chi^2$ by 730.
We now show that, even without detailed modeling, one can infer from 
the form and amplitude of these residuals that the microlens parallax
has amplitude $\pi_\e \ga 0.7$.  Together with the above determination
that $\theta_\e\sim 0.9\,$mas, this implies (even without detailed
modeling) a host mass 
$M_h=\theta_\e/\kappa\pi_\e\la 0.15\,M_\odot$, planet mass 
$m_p=q_2 M_h\la 2\,M_\oplus$,
and distance $D_L = AU/(\theta_\e\pi_\e + \pi_s)\la 1\,$kpc.  Detailed
modeling then confirms these simple by-eye estimates.

The first point is that microlens parallax is a vector 
$\bpi_\e=(\pi_{\e,\parallel},\pi_{\e,\perp})$, with
the direction being that of the lens-source relative proper motion 
$\bmu$.
This is because motion parallel to Earth's instantaneous acceleration
(at the peak of the event) leads to a very different lightcurve distortion
than perpendicular motion.  For parallel motion, the lens ``slows down''
during the event, so the rise toward peak is faster than the fall.  Hence, 
(in the approximation of uniform Earth acceleration) there
is an anti-symmetric distortion, with the data below the non-parallax
model before peak and above after peak.  For perpendicular motion,
by contrast, the distortion is symmetric.  From 
the point-lens magnification formula \cite{einstein36}
$A=(u^2+2)(u^4+4u^2)^{-1/2}$, one can easily work out that the distortion is
given by
\begin{equation}
{\delta A\over A}\simeq 
4 \biggl({a t_\e^2\over\au}\biggr)
G_1(u)\pi_{\e,\parallel}-
\biggl({a t_\e^2\over\au}\biggr)^2
|G_3(u)|\pi_{\e,\perp}^2;
\qquad
G_n(u) = {u^n\over(u^2+2)(u^2+4)},
\label{eqn:parallax}
\end{equation}
where $a$ is Earth's (assumed uniform) acceleration and where we have
assumed $u=(t-t_0)/t_\e$, i.e., $u_0\ll 1$.  Because the target was
at quadrature 35 days before peak, Earth's acceleration can be
treated as roughly constant during the entire pre-peak interval shown
in Fig.~3.  Since the second term in Equation~(\ref{eqn:parallax}) is
quadratic in $\pi_\e$ and has smaller coefficient, it can usually
be ignored to first approximation.  Hence,
$\delta A/A$ is expected to peak at $u=-0.95$ with a value
$\delta A/A=-0.27(a t_E^2/\au)\pi_{\e,\parallel}\rightarrow 0.087\pi_{\e,\parallel}$
where we have used $t_\e=33\,$day and $a=(2\pi/{\rm yr})^2\au$.  From
Fig.~3 the actual peak deviation is $\Delta I=0.063$ at $u=-0.82$,
which corresponds (after accounting for $(F_b/F_s)_{\rm OGLE}=0.25$)
to $\delta A/A= 0.068$ and hence $\pi_{\e,\parallel}=0.78$.  This compares
to the model fit of $\pi_{\e,E}=0.67\pm 0.04$ (noting that
Earth's acceleration is nearly due East at quadrature).  Unfortunately,
only $\pi_{\e,\parallel}$ can be read directly
off the lightcurve: $\pi_{\e,\perp}$ can only be deduced from detailed
modeling.  Nevertheless, this estimate of $\pi_{\e,\parallel}$
places a lower limit on $\pi_\e$ (hence upper limits on the host mass,
planet mass, and distance).

\subsection{Binary Orbital Motion}

The relative transverse velocity of the two binary components can
be measured from two distinct effects.  First, the very sharp
features and high-density coverage of the central caustic allow detection
of subtle changes in the caustic shape and orientation between entrance
and exit due to such transverse motion, even though the interval between
these caustic passages is only of order a day.  For the close binary
models, this is the only source of information.  For the wide binary
models the timing and height of the ``bump'' give the position of
the source relative to the companion roughly 300 days before peak,
and this can be compared to the ``predicted'' position based on the
caustic morphology that is measured at peak.  The model parameters
that capture this effect are the two component vector 
$\bgamma=(\gamma_\parallel,\gamma_\perp)$, which is related to the
physical transverse relative velocity 
$\Delta {\bf v}_\perp = D_L \theta_\e s\bgamma$.  The measurement
of $\bgamma$ can help discriminate between otherwise degenerate models
through $\beta$, the instantaneous ratio of the projected kinetic energy
to the (absolute value of) projected potential energy
\begin{equation}
\beta = \bigg({E_{\rm kin}\over |E_{\rm pot}|}\biggr)_\perp
={(\Delta v_\perp)^2 a_\perp\over 2G M_{\rm tot}}
= {\kappa M_\odot{\rm yr}^2\over 8\pi^2}{\pi_\e s^3\gamma^2\over 
(1+ q_3)\theta_\e(\pi_\e + \pi_S/\theta_\e)^3}.
\label{eqn:beta}
\end{equation}
There are two key points about $\beta$.  First, if the system
is bound, then $\beta$ must strictly satisfy $\beta<1$.  Second,
it is highly improbable (in a sense that we quantify below) that
$\beta\ll 1$.

Before discussing these, we note that there must be only two
stars (plus planet) giving rise to these phenomena, not three stars 
(plus planet)
Naively one might think that the double-horned peak could be due
to a close binary, while the ``bump'' one year earlier was due
to a third star.  However, if there were three such stars, their
combined effect near peak would not be a simple quadrilateral
caustic (See Fig.~2) but a much more complicated self-intersecting
caustic, which would cause multiple entrances and exits in the data.
The fact that the ``double horned peak'' is simple shows that
there are only two stars.

Next, why must the two stars be bound?  That is, why is it not possible
that the source has simply passed by two unrelated stars (one with
a planet) roughly
one year apart?  The rate at which any given source is microlensed
toward these fields is $\Gamma\sim 10^{-5}\,{\rm yr}^{-1}$.  Hence,
the probability for a second encounter, within 1 year and within
3 Einstein radii is $3\times 10^{-5}$.
Even if we take a more generous attitude that what is essential
is the binary-induced central caustic and so the chance projection
of a second star within a $15\,\theta_\e$ circle (rather than the
additional restriction of lying near the source path), a similar
calculation yields $p\sim 2\times 10^{-4}$.  This should be compared
to the $p>10\%$ probability that a given star has a binary companion
within 15 AU.  Hence, this system is very likely to be bound.

We argued in the body of the paper that the system was a wide
binary because there was a ``bump'' in the data very near the time,
amplitude and duration predicted by the wide-binary solutions due to the 
position of the host's companion in those models.  However, there are
several ``structures'' in the baseline data that are due to low-level
correlated noise.  Such structures are often seen in microlensing
data but are ignored because they have no impact on the event analysis.
However, because the ``bump'' plays an important role in the present
case, we must take a closer look.  As discussed in Section 2,
we spent considerable effort tracking down the apparent long-term
brightening of the source.  Hence,
while it would be very strange if the largest and best
defined of these structures happened to basically coincide with the height,
width, and time predicted by the wide model, it is still of interest
to probe the wide/close degeneracy with independent arguments.

The two close models have $\beta=0.00162\pm 0.00105$ and 
$\beta=0.00033\pm 0.00048$.  For simplicity of 
illustration in the following arguments, we choose $\beta=10^{-3}$.
There are three ways in which a binary system can have low $\beta$.
First, the two components could be in a wide circular orbit but are
projected to very close apparent separation.  The probability for this
viewing angle is $p\sim 2\beta^2\rightarrow 2\times 10^{-6}$.  Second they
could be on circular orbits in which the components happen to be
traveling directly along our line of sight (or very close to it).
The probability for this is again $p\sim 2\beta^2\rightarrow 2\times 10^{-6}$.
Third, they could be observed near apocenter in face-on highly eccentric
orbits.  The required eccentricity is then $e\simeq 1-\beta/2=0.9995$.
First, there are no binaries observed in nature with eccentricities
anywhere near this value.  But even if this were one of the few such
systems, it would have a pericenter $q\sim a_\perp\beta/4 <0.1 R_\odot$
which is not physically possible.  Of course, one could imagine combinations
of these possibilities.  For example, if we chose a more reasonable
eccentricity of $e=0.9$, then the pericenter would be physically allowed,
and the required viewing angle less restrictive, but still $p\sim 10^{-5}$.
Hence, very low $\beta$ is highly improbable to the point of being ruled out.

The two wide solutions both have central values $\beta>1$, which would
be unphysical.  For the wide(-) solution, $\beta$ is within $1\,\sigma$
of being physically allowed and within $2\,\sigma$ of having a ``typical''
value.  On the other hand, the wide(+) solution is formally favored by
$\Delta\chi^2=24$.  This proves that there are systematic errors
at the level of $\Delta\chi^2\sim 20$, which is not surprising given
the level of correlated noise at baseline and also given experience
with previous microlensing events.  That is, we can be confident that
a ``bump'' is detected $\sim 300$ days before the main peak, but
we are cautious about deriving detailed parameters based on the
morphology of this bump, which is visible only in binned data.
Therefore, we consider both the wide(+) and wide(-) solutions
to be acceptable.  Since these represent similar physical systems,
this ambiguity does not impact our conclusions.

We note that for many events, the parallax and orbital motion parameters
are correlated, and in particular $\pi_{\e,\perp}$ can be highly correlated
with $\gamma_\perp$ \cite{skowron11} because these two parameters
induce similar distortions on the wings of the lightcurve.  However,
this is not an issue in the present case because while the parallax signal
does indeed come from the wings of the lightcurve (see Fig.~3),
the orbital motion parameters are determined from the location and
height of the ``bump'', which occurred well within the ``baseline'' region
of the lightcurve, long before the onset of the rising wing.

\bibliography{scibib}
\bibliographystyle{Science}

\noindent 
{\bf Acknowledgments}:
We acknowledge:
NSF AST-1103471 (AG,BSG,JCY); NASA NNGX12AB99G (AG,BSG,RWP);
OSU Fellow and NASA Sagan Fellow (JCY);
ERC AdG 246678, IDEAS program (AU);
Polish MNiSW IP2011026771 (SK);
Korea NRF CRIP 2009-0081561 (CH);
US-Israel BSF(AG,DM);
JSPS23340044, JSPS24253004 (TS);
NZ Marsden Fund (PCYK);
Israel SF PBC-ICORE 1829/12 (DM);
NSF grant AST-1211875 (DB).
Data available at \hfil\break
http://astroph.chungbuk.ac.kr/$\sim$cheongho/OB130341/data.html

\end{document}